# Temperature-dependent magnetic particle imaging with multi-harmonic lock-in detection

Thinh Q. Bui, Mark-Alexander Henn, Weston L. Tew, Megan A. Catterton, Solomon I. Woods

T.Q.B., W.L.T, M.A.C, S.I.W are with the National Institute of Standards and Technology, Gaithersburg, MD 20895 USA (e-mail:thinh.bui@nist.gov).

M.A.H is an associate with the National Institute of Standards and Technology, Gaithersburg, MD 20895 USA, and University of Maryland, College Park, MD, USA 20742.

ABSTRACT

Advances in instrumentation and tracer materials are still required to enable sensitive and accurate 3D temperature monitoring by magnetic particle imaging. We have developed a magnetic particle imaging instrument to observe temperature variations using MPI, and discuss resolution dependence on temperature and harmonic number. Furthermore, we present a low noise approach using lock-in detection for temperature measurement, and discuss implications for a new detection modality of MPI.

I. Introduction

Magnetic particle imaging relies on measurement of the magnetization of magnetic nanoparticles (MNP) driven by a strong AC magnetic field [1]-[10]. At NIST, our efforts are directed at establishing magnetic nanoparticles as an SI-traceable nano-thermometer standard for *in situ* thermal imaging, and relying on magnetic particle imaging (MPI) as a core technology. Potential prospects for theranostic applications of temperature sensing [11]-[20] and hyperthermia [16]-[20] based on MPI have been proposed previously. The quantitative nature of thermal imaging with magnetic nanoparticles is demanding and requires thorough investigation of technical noise and drifts in the measurement system. Yet, a systematic evaluation of temperature on the sensitivity and spatial resolution of MPI remains an unresolved problem.

For MPI, the magnetization response of MNPs driven into saturation by a sinusoidal excitation field are detected by inductive coil sensors. Typical MPI systems record the magnetization response in the time domain with a digitizer, and the harmonic components are obtained by digital Fourier transform. While this procedure has the advantage of detecting a broadband harmonic signal, often exceeding 1 MHz, the broad bandwidth carries along with it excess broadband noise. Previously, a narrowband method using intermodulation frequency detection (a single harmonic and its intermodualtion frequency) was introduced to improve MPI noise performance [21]. However, by singling out individual harmonics, high harmonic information content that is responsible for high imaging spatial resolution is not available, commonly described as a resolution-bandwidth trade-off [22]. Instead of purposefully reducing bandwidth, it was later proposed that low noise performance can be achieved by careful noise-matching inductive coil sensors to pre-amplifiers with a reported 11x gain in noise performance [23],24], but this approach requires tailor-made transformer-like networks specific to each MPI system. An additional problem with this strategy is that the complex impedance of the inductive coil introduces a frequency-dependence of noise that complicates noise matching over too wide of a bandwidth > 1 MHz.

Here, we propose an alternative and more general approach by implementing a massively-parallel lock-in detection scheme in which many harmonics are measured simultaneously while still leveraging the narrowband detection of the lock-in technique, resulting in a reduction in measurement noise by a factor of 100-1000, depending on the desired acquisition speed. In contrast to [21], we are detecting many harmonics from the particle response in a narrowband fashion, thereby overcoming the resolution-bandwidth trade-off. Lock-in detection can also reduce the data streaming requirements, since it does not require saving the full time response. It is a general goal of magnetic imaging to reach the body noise limit, which can potentially be achieved at a high frequency (MHz range) by detecting at a high harmonic number and reaching coil resistance (thermal) noise. In this work, we present our multi-harmonic lock-in detection approach, characterize noise performance, and compare performance of parallel lock-in detection against the commonly used broadband detection with a digitizer.

In addition to concentration imaging, nanoparticle thermometry has been a longstanding goal of MPI. Two key approaches for nanoparticle thermometry have been proposed: multi-color, system matrix-based [13],[25],[26] and multi-harmonic-based measurements [11],[14],[15],27]. The system matrix approach requires measurements of calibration system matrices to reconstruct both concentration and temperature, and a linear interpolation step for intermediate temperatures [13]. In this work we will focus on the multi-harmonic technique, where the temperature can be determined by measurement of two harmonics and their ratios. Here, we will also present temperature-dependent MPI, which relies on the multi-harmonic detection, and show that performance is improved by lock-in detection. Finally, we discuss the observed spatial resolution dependence on harmonic number, which will have important implications for image reconstruction when temperature is involved.

**II. MPI instrumentation**

The MPI instrumentation developed at NIST was designed for thermal imaging and for thermometric metrology using magnetic nanoparticles. Inspired by our previous temperature-tunable AC magnetometers [28],[29], the MNP sample holder is machined out of thermally conductive Shapal ceramic. The temperature of the sample is tuned by water-cooling with a recirculating chiller, and the temperature is measured by two 100 ohm platinum resistance thermometers. An additional water-cooling line is used to control the temperature of the excitation solenoid coil to ensure thermal stability over long acquisition times. The selection field (gradient field) is generated from a pair of permanent magnets (diameter = 50 mm, height = 50 mm, N52 grade NdFeB) separated by ≈30-40 mm to produce the field-free-point (FFP). The FFP is fixed in our MPI imager, and spatial scanning is achieved by mechanical translation of a 3D stage in which the sample holder and excitation coils are mounted. This image acquisition scheme relies on spatial mechanical step-scanning at a maximum rate of ≈5 Hz, which makes it more similar to the x-space method [8],[9],[30],[31] that uses low frequency (≈10's Hz) shift coils rather than the original Lissajous pattern-based electromagnetic shifting with three orthogonal fields near 25 kHz [1],[6],[32].The transmit chain is a resonant circuit at the excitation frequency $f_0$ = 31.75 kHz with a second order low-pass filter to attenuate (>50 dB) the higher harmonic (feedthrough) signals. Magnetization response from AC excitation (10 mT) is recorded with an inductive receive coil, wound as a first-order gradiometer in series with an additional $\approx-30$ dB notch filter at the excitation frequency. This output is directly connected to a SR560 pre-amplifier, subsequently connected in series to an impedance matched, cascaded notch filter stage with ≈60 dB of attenuation at the $f_0$ and ≈40 dB at $2f_0$. Output from the second notch filter is sent to a final low-pass filter stage (SRS SIM965). Finally, the filtered magnetization response is directed to either a digitizer (GaGe, 16 bit, 1 GSa/s) or lock-in amplifier (Zurich HF2LI) to measure the harmonic amplitudes for image reconstruction. For the digitizer, the harmonic amplitudes are determined from the Fourier transform of the time domain signal. The multi-frequency capability of the HF2LI lock-in amplifier allows the recording of six independent harmonics simultaneously in a parallel fashion. For quantitative imaging, we use a glass phantom with dimensions of 8 x 12 mm (diameter x height). The phantom has four channels (1 mm diameter, 8 mm height) where MNP samples can be introduced and sealed by epoxy. To demonstrate the 3D imaging capability of our MPI imager, Fig 1 shows the 3D image of this glass phantom in two, rotated orientations, with four clearly resolved channels. For a more systematic evaluation of spatial resolution, we used a 3D printed acrylic phantom shown in Fig. 2. Here, the sets of four channels have different edge-to-edge hole spacing to evaluate the spatial resolution using Rayleigh criterion. The 2D image in Fig. 2 reveal that we can resolve 0.25 mm in the x-direction, the direction of the highest gradient field. These results, along with the measured point-spread-function (PSF) of Vivotrax, indicate that the magnitude of the gradient field is ≈20 T/m for the x-axis and ≈10 T/m for the y-axis. These results are consistent with our COMSOL simulations of the permanent magnet pair.

*A. Parallel harmonic lock-in MPI*

A voltage-based measurement system of MPI comprised of an inductive coil sensor connected to a pre-amp has SNR given by [21],[23]:

$$SNR \propto \frac{S \cdot \frac{dm}{dt}}{\sqrt{\left(4k_B T R_L + \overline{e_n^2} + \overline{\iota_n^2}\left|Z_L^2\right|\right) \cdot \Delta f}} \quad (1)$$

The quantity in the numerator is the product of the coil sensitivity, $S$, and voltage induced by the magnetic nanoparticles, *dm/dt*. The denominator denotes the noise sources from the sensor, an inductive coil sensor with real ($R_L$) and complex impedance ($Z_L$), and the pre-amp. The pre-amplifier has a voltage and current noise spectral density (NSD,$V/\sqrt{Hz}$), $\overline{e_n}$ and $\overline{\iota_n}$, respectively. Since Δf is the detection bandwidth, the best SNR is obtained by minimizing the bandwidth. This ideal scenario is, however, at odds with the MPI technique in that it is desirable to record all detectable information content-carrying harmonics. Reducing the sampling rate (detection bandwidth) in the digitizer system could improve SNR, but at the expense of spatial resolution [22]. Electronic noise matching in the receive chain has been proposed, but the complex impedance of the inductive coil introduces a frequency dependence that complicates noise matching strategies over broadband frequencies [21],[23],[32]. NSD measurements for our signal chain (IC with 2nd order notch filter (Fig. 14, SI)) does indeed show a complex frequency dependence with increasing noise density at higher (>1 MHz) frequencies. To overcome the noise matching and bandwidth requirement incompatibility challenges, we present a lock-in detection method with parallel multi-harmonic capabilities that can improve SNR by permitting detection at high frequencies without the need for broadband noise matching.

Relying on measurement of only a few harmonics has been proposed for in-situ temperature measurement [11],[15], and also as a general modality for imaging [9],[24]. Isolated harmonics are most sensitively measured with lock-in detection, where the noise bandwidth (1 to 100 Hz range) is most ideal and adjustable based on the desired acquisition speed. With a digitizer, a slower acquisition rate will reduce broadband noise, but also reduce the detection bandwidth to observe the rich harmonic spectrum of saturated MNPs, which often exceed 1 MHz. For detection of a single harmonic, the equivalent noise bandwidth $\Delta f_{lock-in}$ for lock-in detection is set by the lock-in time constant, T, which typically in our experiment is 10 ms. For a first order RC filter, $\Delta f_{lock-in} = \frac{1}{4T}$, which gives a $\Delta f_{lock-in} \approx$ 25 Hz. Compared with a 1 MHz bandwidth digitizer acquisition, the SNR enhancement of lock-in detection relative to digitizer is $\sqrt{10^6/25} \approx 200$. This analysis shows that multi-harmonic lock-in measurement can reduce broadband noise associated with the inductive coil detection. Improving MPI noise performance by electronic noise matching in the receive chain [23],[24],[33],[34] can only further improve overall performance.

*B. Noise performance and SNR of lock-in detection*

Despite the superior noise performance of lock-in detection, commercial lock-in amplifiers often limit to one or two harmonics at a time. Yet, if many of the particle harmonics could be measured in a parallel fashion, then SNR is optimized without sacrificing spatial information content by filtering higher harmonics. These considerations motivated us to compare the noise and imaging performance of multi-harmonic detection and time-domain detection with a digitizer. First, we compared the detection background noise by analyzing the time domain signal when the receive chain is connected to either the digitizer or the lock-in amplifier (Fig 13 (SI)). Here, the sampling conditions are set the same for both instruments, which provides a direct comparison of the analog-to-digital conversion (ADC) noise of the data acquisition devices. As presented in Fig 13 (SI), the RMS noise ($V_{RMS}$) of the lock-in amplifier signal is $\approx$ 3x lower noise than the digitizer for the same acquisition rate

To resolve the different contributions to the total noise, the noise spectral densities were measured for (1) the lock-in amplifier only and (2) the IC with the notch filter only, and (3) all components in the receive chain including the pre-amp, as shown in Figure 14 (SI). The 1 MHz bandwidth of the pre-amp suppresses noise at above 1 MHz, but increases the noise level by a factor of 100 in the range of interest of 30 kHz to 1 MHz. Fig. 14c shows a higher noise-baseline from broadband detection of 1 MHz. These NSD measurements in Fig. 14b and c were used to calculate the total RMS noise by taking the integral of $NSD^2$ over the full 1 MHz bandwidth. This calculation agrees with the measured RMS noise in Fig. 13 (SI).

To validate the enhanced noise performance of multi-harmonic lock-in detection with real samples, we compared images of the Vivotrax glass phantom used in Fig. 1 acquired with the lock-in amplifier and with the digitizer. Figure 3 show the 1D image slice across two channels of the glass phantom. The four panels show a comparison of the 3rd to 9th harmonic signals from the lock-in and digitizer. The SNR enhancement is most clearly observed in the 7th harmonic, with an improvement of $\approx$ 3x with the lock-in relative to the digitizer, in reasonable agreement with the predicted noise improvement of $\approx$ 200x from broadband noise reduction from 1 MHz (digitizer) to 25 Hz (lock-in).

*C. Temperature dependent MPI*

Temperature impacts the equilibrium magnetization as well as the dynamics of magnetization, which will be imprinted on the MPI signal. There have been a number of reported studies on the temperature dependence of suspended magnetic nanoparticles [11]-[15],[27],[35]-[38], yet quantitative effects of temperature on MPI signal and spatial resolution is not fully understood. Using the imaging phantom in Fig. 1, we measured the temperature dependence of Vivotrax to assess temperature's impact on image reconstruction with MPI.

The temperature of the Vivotrax sample was tuned from 278 K to 305 K and allowed to equilibrate to $\pm$ 50 mK prior to imaging. The 2D MPI images for 4 of the 7 temperature steps are plotted in Fig. 4 as columns, with each 2D image being an average of 3 repeated measurements. It is evident by visual inspection that the 9th harmonic has a stronger temperature dependence than the 7th harmonic. The quantitative dependence is obtained from the 1D representation of Fig. 4 versus pixel number, and is shown in the Fig. 15 (SI) top row plots. For each plot, the overlay of 7 temperatures is shown. To obtain the magnitude, the integrated areas are calculated for the peaks labeled in Figs. 15a-b (SI) and are presented in Figs. 15c-d (SI) . The slopes for each harmonic gives the temperature sensitivity [28]. Since each peak gives a different spatial location of magnetic nanoparticles, the agreement in their calculated sensitivities gives evidence for the stability of the system over long periods of scanning and averaging (30 minutes of acquisition time at each temperature, 3 averages). To gauge the short and long time stability, the Allan variance plots for a single image pixel are displayed in Fig. 16 (SI). In our instrument, statistical noise can be improved by averaging up to 10 seconds before systematic drifts degraded noise performance.

Previous temperature-dependent MPS measurement of Vivotrax samples in an 8 mm diameter sapphire vial show a negative temperature dependence on harmonic amplitude predicted by the Langevin function [28], which we reproduced with the current MPS instrument [29]. In contrast to the 8 mm diameter glass vial used, our MPI results for Vivotrax using a glass phantom as a sample vessel display the opposite (positive) temperature dependence. One possible explanation for this is that the narrow (1 mm diameter) channel of the phantom induces MNP aggregation, which hinders Brownian motion. If Brownian motion is blocked and the temperature dependence is dominated by Néel rotation, the magnetization response displays a positive temperature dependence as reported in the literature from MPS [37] and MPI [38] studies. We independently verified with MPS that the phantom vessel displays a positive temperature dependence consistent with our MPI results, while the larger glass vial shows the opposite trend. Conversely, measurement of Vivotrax in the larger glass vial using both the MPS and MPI instrumentation showed the same (negative) temperature dependence (Fig. 17, SI).

The imaging spatial resolution dependence on harmonic number and temperature were also measured for Vivotrax with the glass phantom. 1-D scans across the two channels of the phantom are presented in Fig. 5 for both the 7th and 9th harmonics. Only a modest change in the FWHM of the peaks was observed. A more quantitative measurement of the temperature dependence is determined from the PSF of Vivotrax (Fig. 6). In contrast to measurements in Fig. 5, the PSF in Fig. 6 includes all harmonics and therefore exhibits significantly higher SNR. By fitting the PSF curves to a Voigt function, the temperature dependence of the FWHM was obtained (Fig. 6b). A modest slope of 0.041(6) mT/K is consistent with a small temperature dependence shown in Fig. 5.

For thermal imaging applications using magnetic nanoparticles, measurement of harmonic ratios, rather than individual harmonic amplitudes, has been proposed to remove the concentration dependence [11]. We recorded 2D images using individual harmonics and also their ratios, and the data is presented in Fig. 7. In the case of harmonic ratios, 5th/3rd and 7th/3rd, the higher harmonic dominates the spatial resolution of the ratio image, which mostly resembles the image of the higher harmonic number. In general, images obtained with either individual harmonics or harmonic ratios clearly show a harmonic dependence of spatial resolution, and suggests a corresponding point-spread-function (PSF) dependence on harmonic number. This concept of a harmonic PSF is crucial for this imaging modality, and will be discussed in detail in the following sections.

**III. Harmonic dependence of spatial resolution**

Previous MPI implementations typically utilize the full time signal for image reconstruction, which obscures the impact of individual harmonics on spatial resolution. The effect of harmonics on the reconstructed image has been introduced previously [2],[31], mainly in the context of feedthrough (1st harmonic) suppression. More importantly, these studies reveal a direct mapping of harmonic frequency space to spatial frequency in the image space for sinusoidal excitation. Our work extends upon these analyses by experimentally determining the impact of individual harmonics on image reconstruction, a new concept denoted as the harmonic PSF. To demonstrate the existence of a harmonic PSF, we acquired MPI images reconstructed using individual harmonics by means of (1) the spectrum of the digitized time signal and (2) the lock-in amplifier demodulated signal. Using both approaches, the apparent dependence of resolution on harmonic number was observed for two different MNP samples. Here, a glass phantom with four channels was used as a sample holder for 70 nm Synomag (Fig. 8) and for Vivotrax (Fig. 9). Figure 8a shows a strong resolution dependence on harmonic number, as the four wells become more spatially resolved with increasing harmonic number. To quantitatively analyze the resolution dependence, an acrylic phantom with two sample wells spaced by 0.5 mm was machined and imaged in 1D (Fig. 8b). The spatial resolution was determined by fitting the FWHM of the measured profile, and the result is plotted in Fig. 8c. The resolution enhancement experiences an exponential improvement at low harmonic number and saturates at higher harmonics. A similar analysis was done for Vivotrax tracers (Fig. 9) with an observed exponential-like dependence of resolution on harmonic number.

*A. PSF measurement*

The spatial resolution of MPI is dependent on the PSF of the magnetic nanoparticles and the gradient of the selection field [39]. By applying a slow varying (10 Hz) bias field at 80 mT$_{pp}$ superimposed on the excitation field, the PSF can be determined in a similar fashion to [40]. Different from this PSF reconstruction procedure that includes all harmonics similar to previous reports, we extended this concept to measure the PSF of individual harmonics. The PSF of the individual harmonics were resolved by two different methods: (1) short time Fourier transform (STFT) of the digitized time response signal and (2) time-swept lock-in detection. For these measurements, the bias field is driven by a triangle waveform with $f_{bias} = 10$ Hz with an amplitude of 40 mT$_p$.

For method (1) using the time signal $f(t)$ from the digitizer, the short time Fourier transform is implemented by first binning the digitized data in defined window segments $w(t)$. At each time shift $t$, a new window $w(t' - t)$ is applied and the FT is calculated of $f(t')w(t' - t)$:

$$\hat{f}(t, u) = \int_{-\infty}^{\infty} f(t')w(t' - t)e^{-i2\pi t'u} dt' \quad (2)$$

This procedure produces a spectrogram, or a plot of the Fourier spectrum at each time segment window. With the bias field turned on, the harmonic spectrum will only appear when the bias field amplitude is near zero, that is when the MNP are not fully saturated. For method (2), the harmonic PSF is obtained with the lock-in by implementing a time-domain acquisition of the output at the specified demodulation frequency ($f_0$, $3f_0$, $5f_0$, etc.) to resolve the time-swept PSF at each harmonic, similar to data in Fig. 6a.

Figure 10 is a step-by-step layout of the STFT procedure. Harmonic PSF data taken with the digitizer (method 1) and lock-in amplifier (method 2) are plotted in Figs. 11a-b. The general shape of the PSF obtained using the two independent approaches display qualitative agreement for each harmonic, and provides experimental justification for the origin of the harmonic dependence of PSF. By inspection, it can be seen that the harmonic PSF displays a decreasing FWHM (higher spatial resolution) with increasing harmonic number, and the FWHM also converges to an asymptotic value. This trend explains our imaging results for nanoparticle samples in Figs. 8 and 9 and the exponential dependence in Figure 8c.

*B. Simulation of harmonic dependence of resolution*

To obtain further validation and insight into the origin of the observed spatial resolution dependence on harmonic number, we performed the following simulation study. The magnetization of a single magnetic nanoparticle is modeled using the Langevin function, i.e.,

$$\vec{M}(\vec{H}) = \mathcal{L}(k\vec{H}) \frac{\vec{H}}{||\vec{H}||}, \text{ with } k = \frac{\mu_0 m}{k_B T}, \quad (3)$$

and $\mu_0$ being the vacuum permeability, $m$ the particle magnetic moment. The moment is a function of the saturation magnetization $M_s$, the Boltzmann constant $k_B$, and the temperature $T$. The PSF $\vec{S}(t)$ is derived by calculating the time-derivative of the Langevin function, Eq. (3) [30]. Finally, the harmonic dependence of PSF is calculated in the same fashion as the STFT and lock-in approaches. Measured and simulated harmonic PSF presented in Figs. 11a-b and Figs. 11c-d, respectively, show that the PSF FWHM, defined by the broad envelope at each harmonic, is narrower with increasing harmonic number, with nearly an exponential-like drop off in resolution [41]. The simulations also reveal the origin of the narrower PSF for higher harmonics: the amplitude decreases faster for higher harmonics relative to lower harmonics for nanoparticles that are increasingly farther away from the FFP region.

We also simulated the temperature dependence of the PSF to validate the observations in Fig. 6. Note that due to the explicit temperature dependence of $k$ in Eq. 3, the magnetization of a single particle shows a corresponding temperature dependence, even if the saturation magnetization was set to be constant. Figure 12 shows the simulated temperature dependence of PSF over the range 250 K to 300 K, in which only a modest change in the PSF FWHM with temperature (0.043 mT/K) was observed, in good agreement with the Vivotrax data from Fig. 6.

**Conclusion**

Towards the goal of advancing temperature measurement and thermal imaging with magnetic nanoparticles, this work comprised experimental and theoretical studies on the harmonic dependence of MPI and measurements of noise performance based on a digitizer and multi-channel lock-in amplifier. The study shows the advantage of using parallel multi-harmonic lock-in detection as a general data acquisition method with exceptional noise performance and straightforward implementation. However, scaling up lock-in amplifiers to multiple independent channels is expensive with commercial solutions. A more cost-effective solution to increase the number of independent lock-in detection channels is to implement high frequency lock-in with FPGA [42]. This will require developments of lock-in electronics that are both noise-matched to the receive chain and operate at high frequencies.

## Appendix

A. Numerical Details

Note, that for this study we assumed all quantities from Eq. 3 to be scalars. With the time-dependent PSF $S(t)$ evaluated at $N+1$ different times $t_{-N/2}, t_{-N/2-1}, \dots, t_0, \dots, t_{N/2+1}, t_{N/2}$, with center point $t_0$, that fall within the window segment $w(t_0) = [t_{-N/2}, t_{N/2}]$, and a triangular shaped window function $W(t)$ that has its maximum at $t_0$, we can calculate the PSF's $k$-th harmonic through the discrete STFT:

$$c_k = \frac{1}{N+1} \sum_{i=-N/2}^{N/2} W(t_i) S(t_i) \exp(-i 2\pi k f_0 t_i). \quad (4)$$

The lock-in measurement for the $k$-th harmonic is simulated by first multiplying the time-dependent PSF $S(t)$ with a reference signal such that:

$$\hat{S}(t) = S(t) \exp(-i 2\pi k f_0 t), \quad (5)$$

and then taking the discrete Fourier transform of Eq. 5 over a time window $w(t)$.

The time resolution in all our simulations was set to $10^{-9}$ s to find a balance between necessary accuracy and memory constraints.

## Acknowledgment

We acknowledge Jo Wu for assistance with the glass imaging phantom. We thank Angela Hight Walker, Cindi Dennis, Michael Donahue, Adam Biacchi, and Frank Abel for helpful discussions. We acknowledge financial support from the NIST Innovation in Measurement Science (IMS) program. Reference is made to commercial products to adequately specify the experimental procedures involved.  Such identification does not imply recommendation or endorsement by the National Institute of Standards and Technology, nor does it imply that these products are the best for the purpose specified.

**FIGURES**

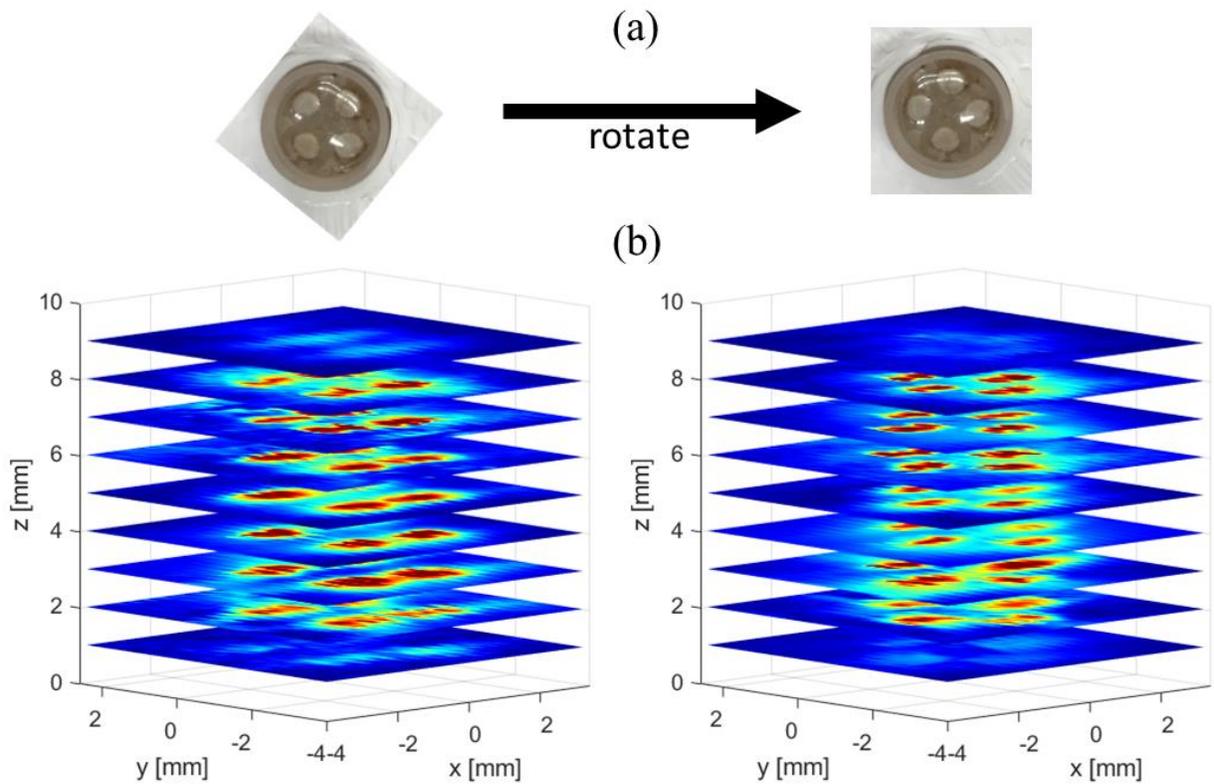

Figure 1: Magnetic particle imaging: a) 4-channel glass phantom containing undiluted (5 mg Fe/ml) Vivotrax tracer at two rotated orientations shown from the top view. The four white circles (diameter = 1 mm, height = 8 cm) are the sample channels. b) 3D-MPI measurement of the phantom, measured by mechanically scanning voxels in steps of 0.25 x 0.25 x 1 mm to map out the 3D phantom image at the two orientations.

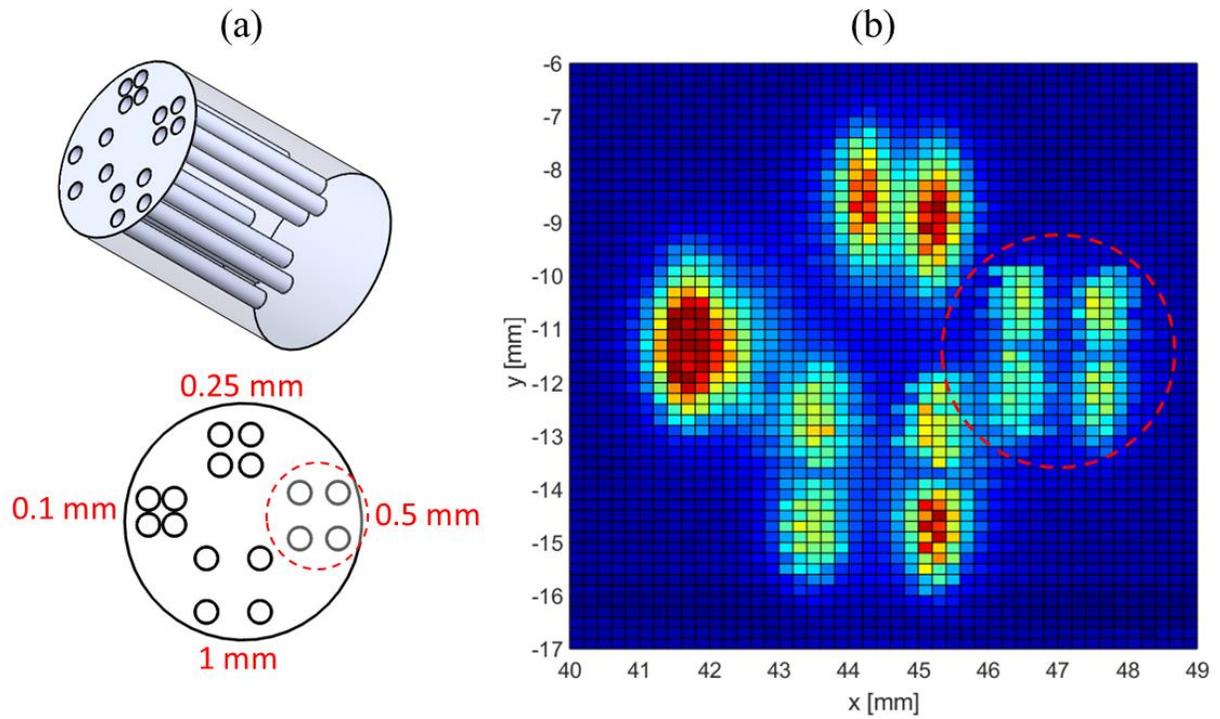

Figure 2:(a) 3D-printed acrylic spatial resolution phantom with 16 channels. The diameter of each channel is 0.75 mm. The group of 4 channels have edge-to-edge separations indicated by the labels on the top view cartoon. (b) 2D MPI image. For clarity, the dashed red circle denotes the image corresponding to the circled portion in the top view cartoon (0.5 mm separation).

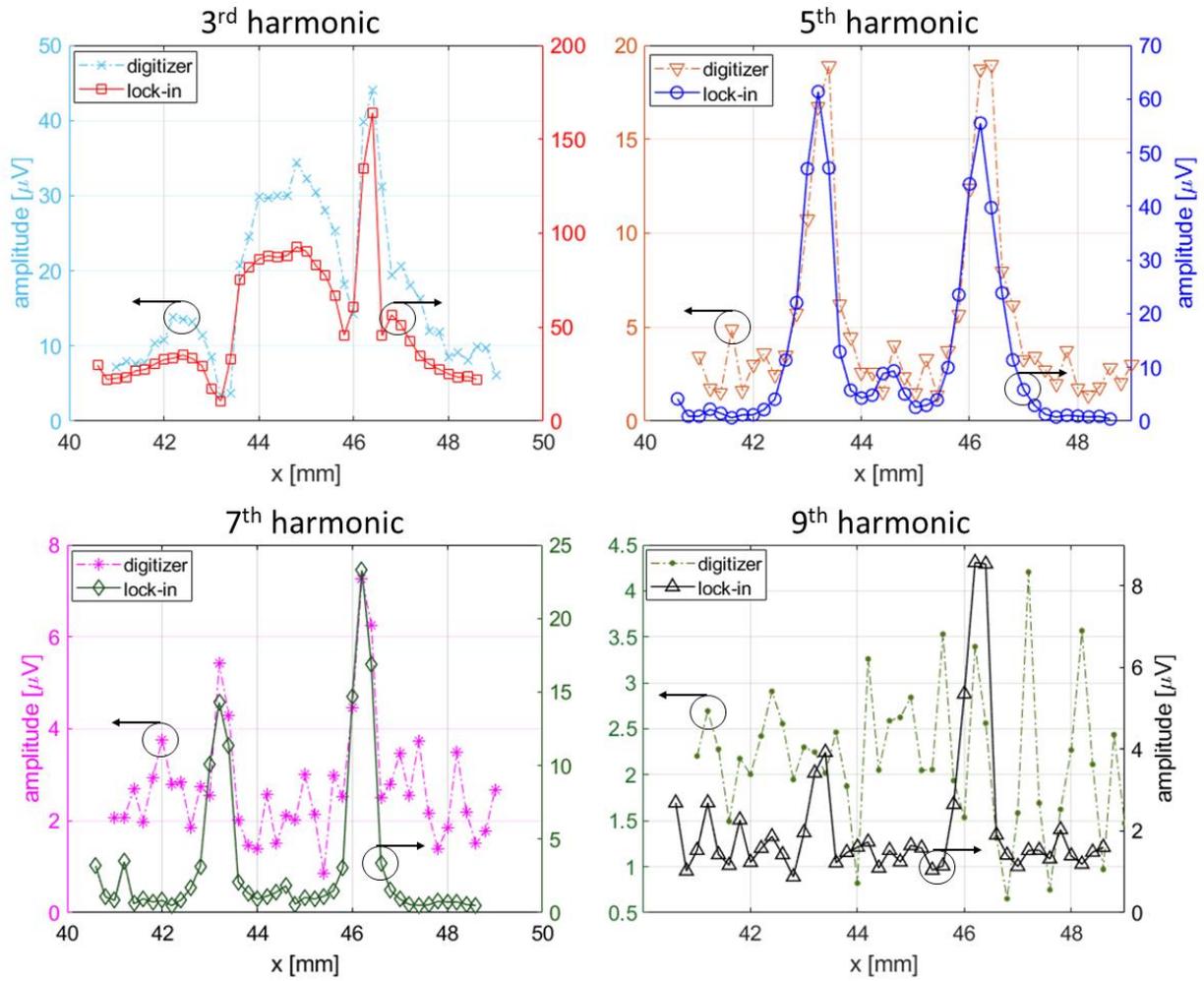

Figure 3: 1D scan of Vivotrax tracer in the glass phantom from Fig. 1. The two peaks correspond to two of the four sample channels. The pre-amplifier gain is set at 10x to avoid the quantization noise floor of the digitizer.

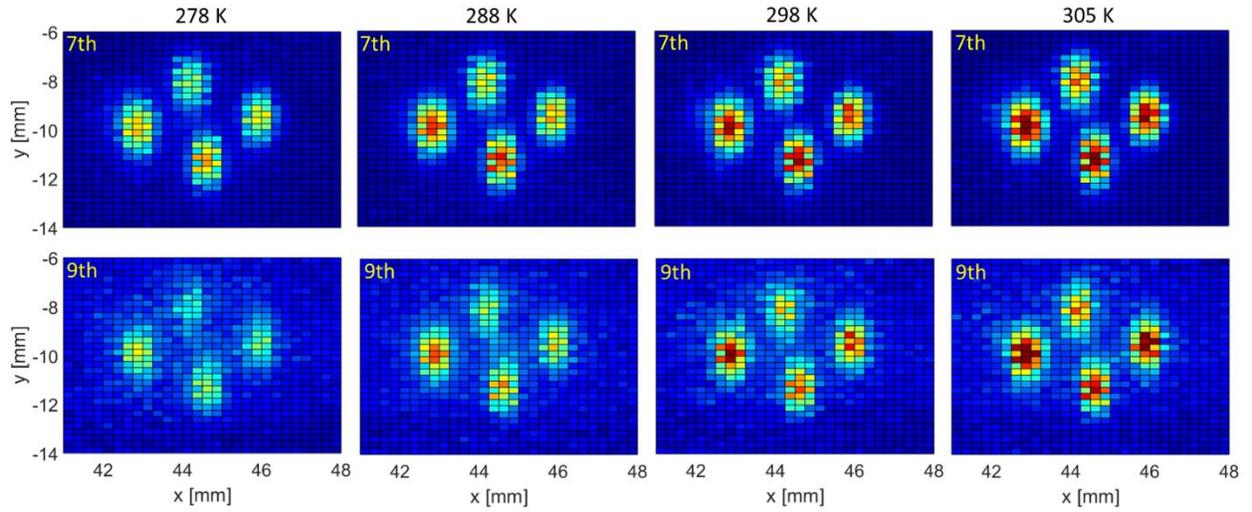

Figure 4: Thermal imaging of Vivotrax in a glass phantom as a function of harmonic number. Each column is a different temperature and each row a different harmonic number.

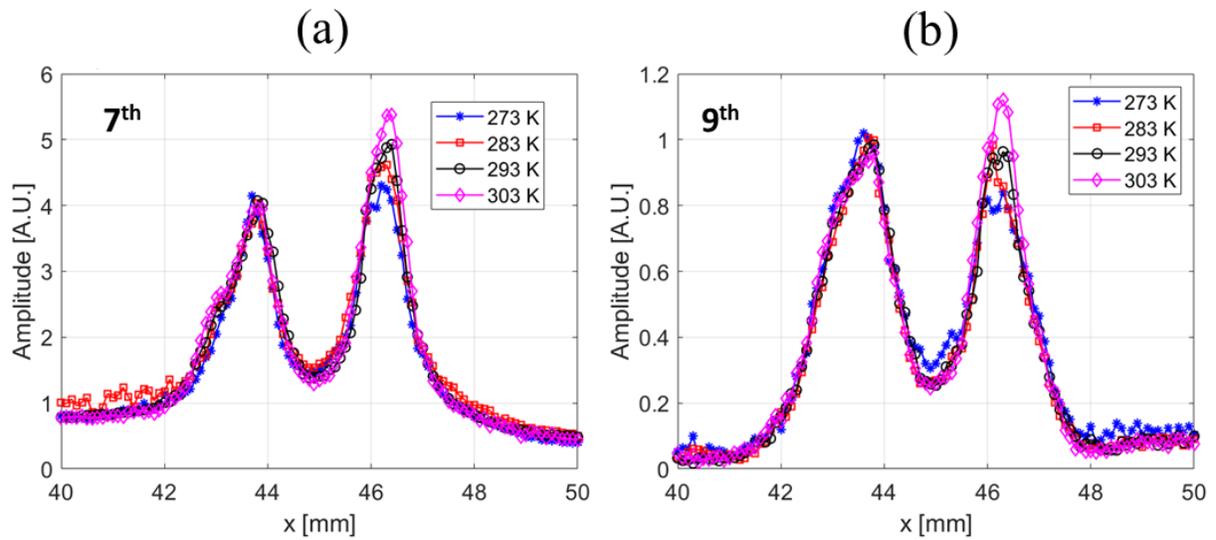

Figure 5: 1D images of Vivotrax as a function of temperature for the 7th (a) and 9th (b) harmonics.

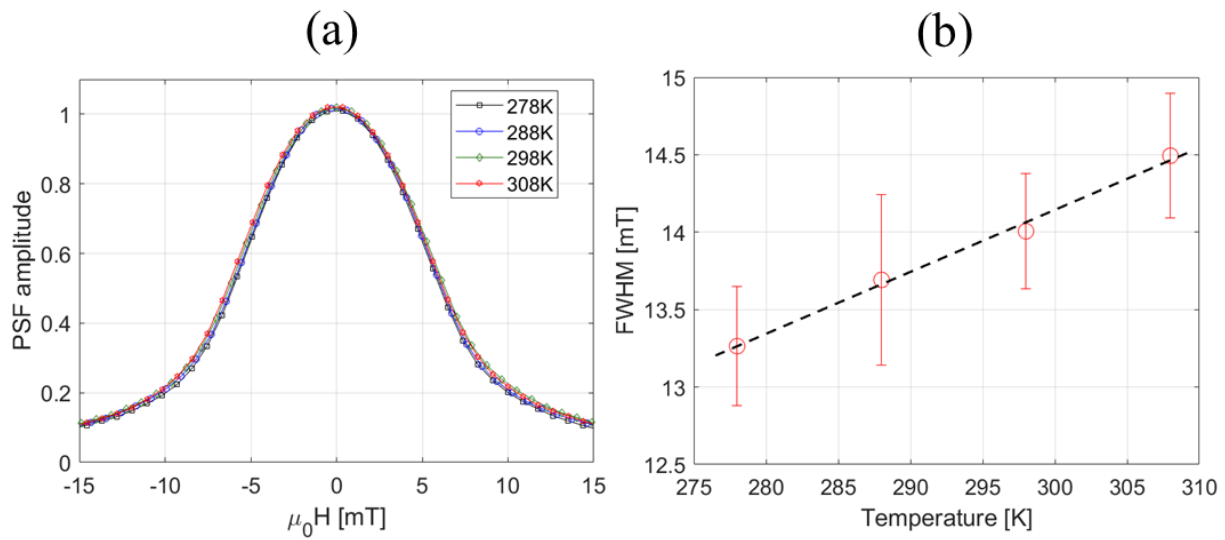

Figure 6: (a) Vivotrax PSF and (b) fitted FWHM as a function of temperature (slope = 0.041(6) mT/K). This PSF includes information from all harmonics.

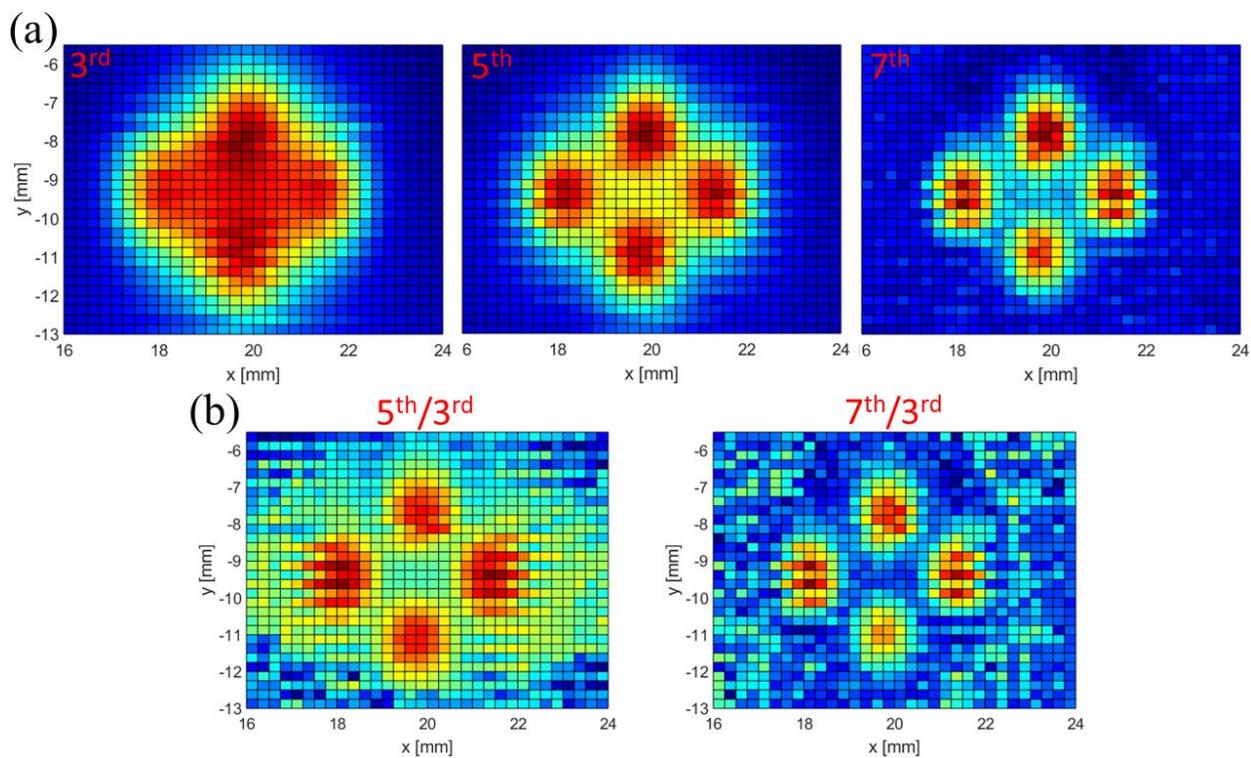

Figure 7: Spatial resolution dependence on (a) harmonic number and (b) harmonic ratio for 70 nm Synomag nanoparticles.

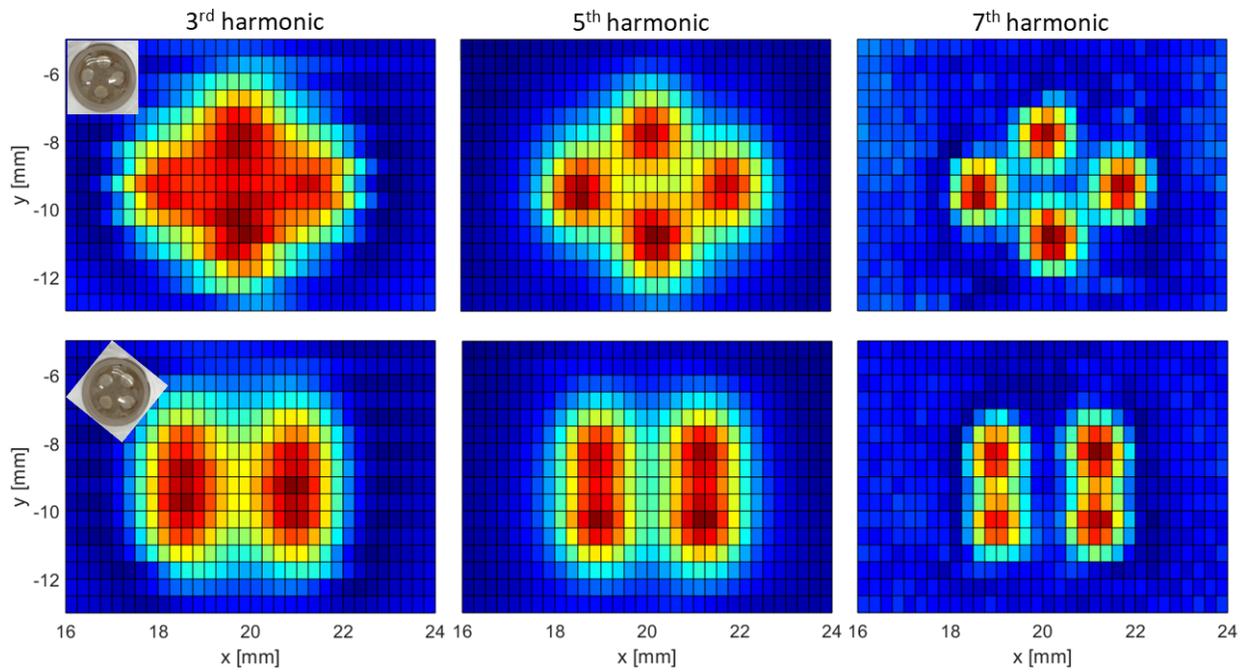

Figure 8: (a) 2D-MPI images of undiluted (15 mg/ml) 70 nm Synomag nanoparticles in a glass phantom. The apparent spatial resolution improves with increasing harmonic number. (b) An acrylic phantom (inset) is used to obtain the 1D image for quantifying the spatial resolution dependence on harmonic number. (c) Fitted FWHM from (b) along with a bi-exponential fit (dashed blue curve) to the FWHM.

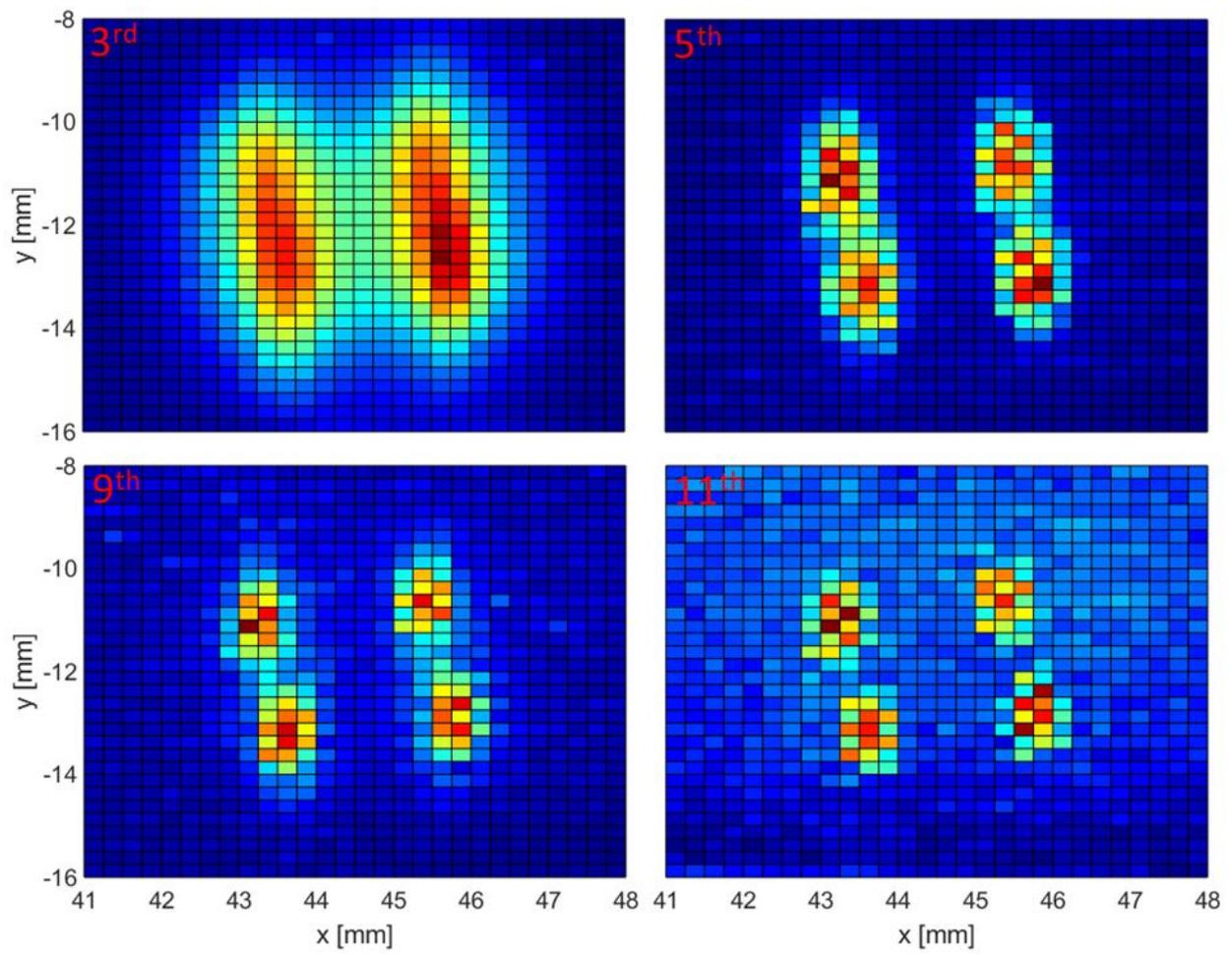

Figure 9: 2D-MPI images of undiluted (5 mg Fe/ml) of Vivotrax tracers in a glass phantom. Vivotrax shows a more gradual harmonic dependence of resolution than 70 nm Synomag.

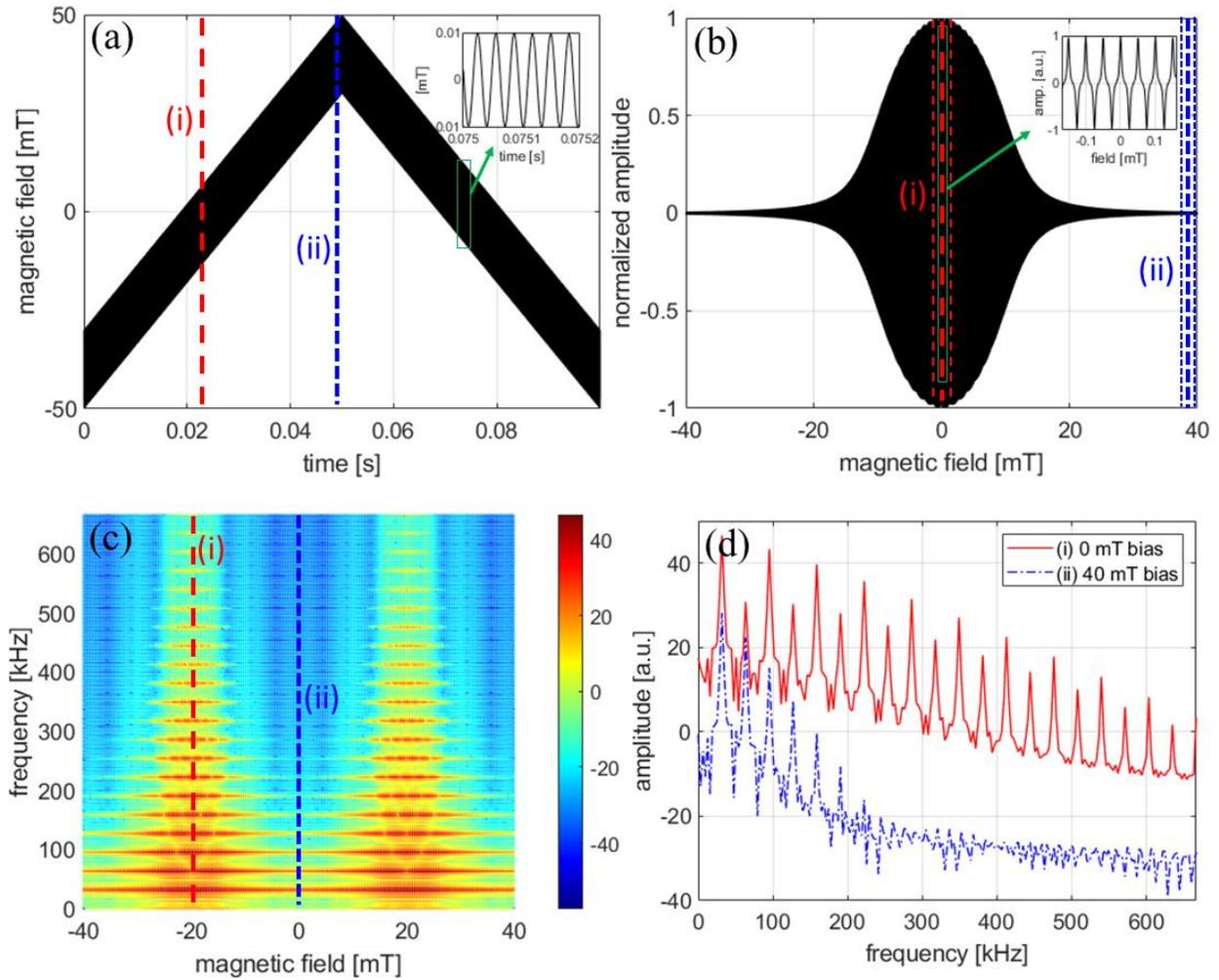

Figure 10: Short time Fourier transform (STFT) procedure for obtaining the harmonic PSF at excitation and bias field frequencies $f_0 = 31.75$ kHz and $f_{bias} = 10$ Hz, respectively. (a) The applied field (sum of excitation and bias) as a function of time. The red line i) denotes the time when the bias field is zero, and the black line ii) denotes the time when the bias field is at its maximum of 40 $mT_p$, (b) zoomed-in view of position i) at the local maximum (thick red line) with the integration time window (thin red lines) and ii) at the the local minimum (thick black line) with the integration time window (thin black lines). (c) Spectrogram of nanoparticle response signal using a triangular shaped window function that results in a 50% overlap in the integration domain. The red and black lines correspond to the time positions i) and ii), respectively. (d) Fourier spectrum for the two different time position i) and ii). The blue vertical lines denote the odd harmonics.

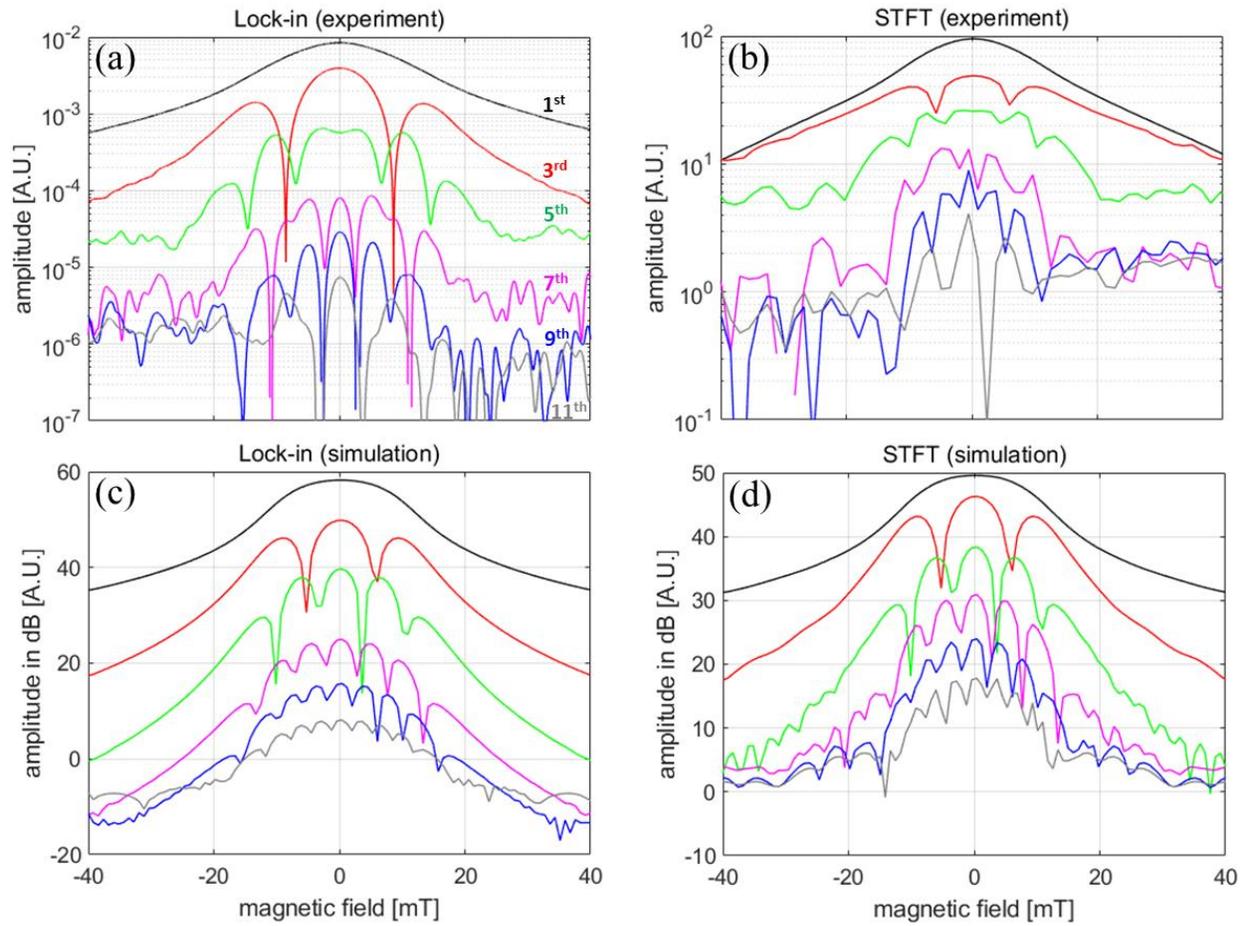

Figure 11: Harmonic PSF measurement of Vivotrax using a digitizer (a,c) and a lock-in amplifier (b, d) at excitation frequency $f_0$ = 31.75 kHz. (a,b) and (c,d) are the measurements and simulations, respectively.

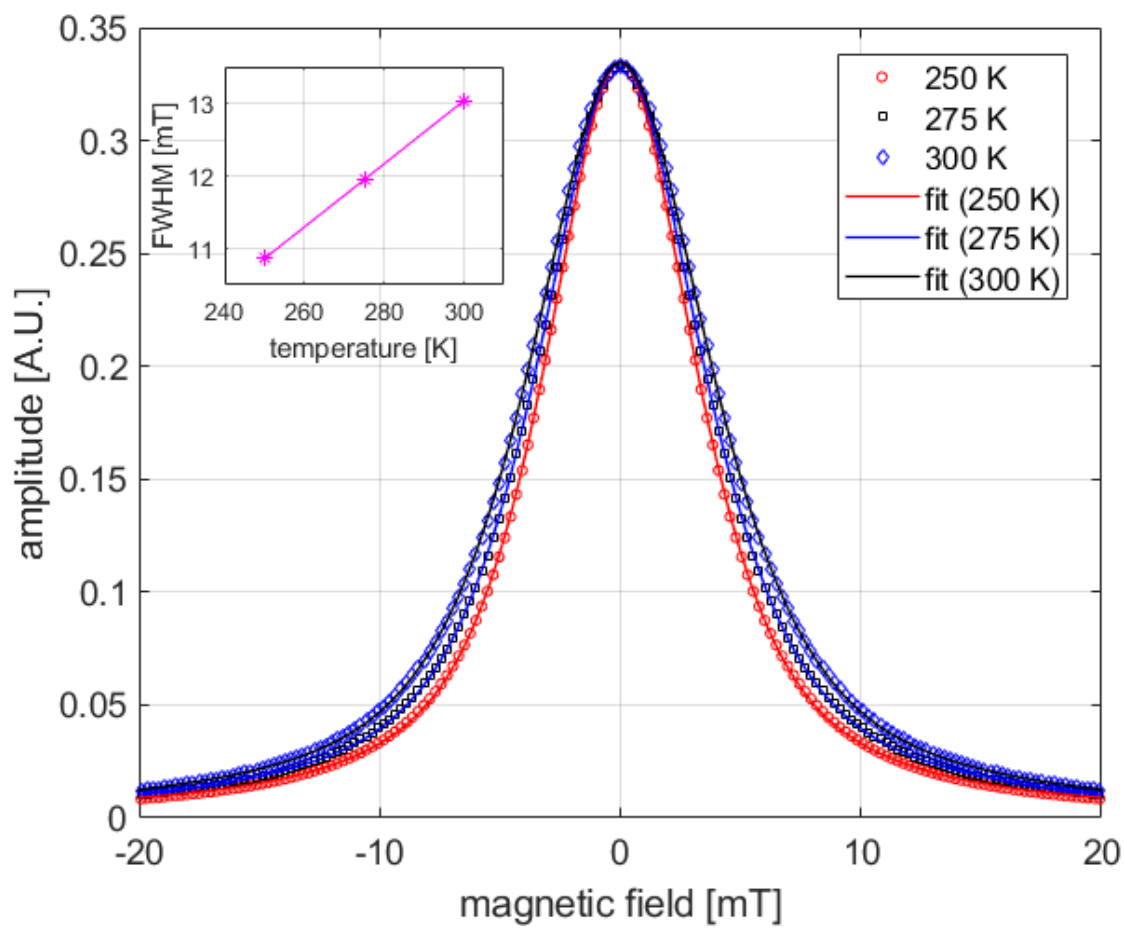

Figure 12: Simulated PSF for Vivotrax at three different temperatures, along with the fits using a Voigt function. The inset shows the fitted FWHM with a slope of 0.043 mT/K.

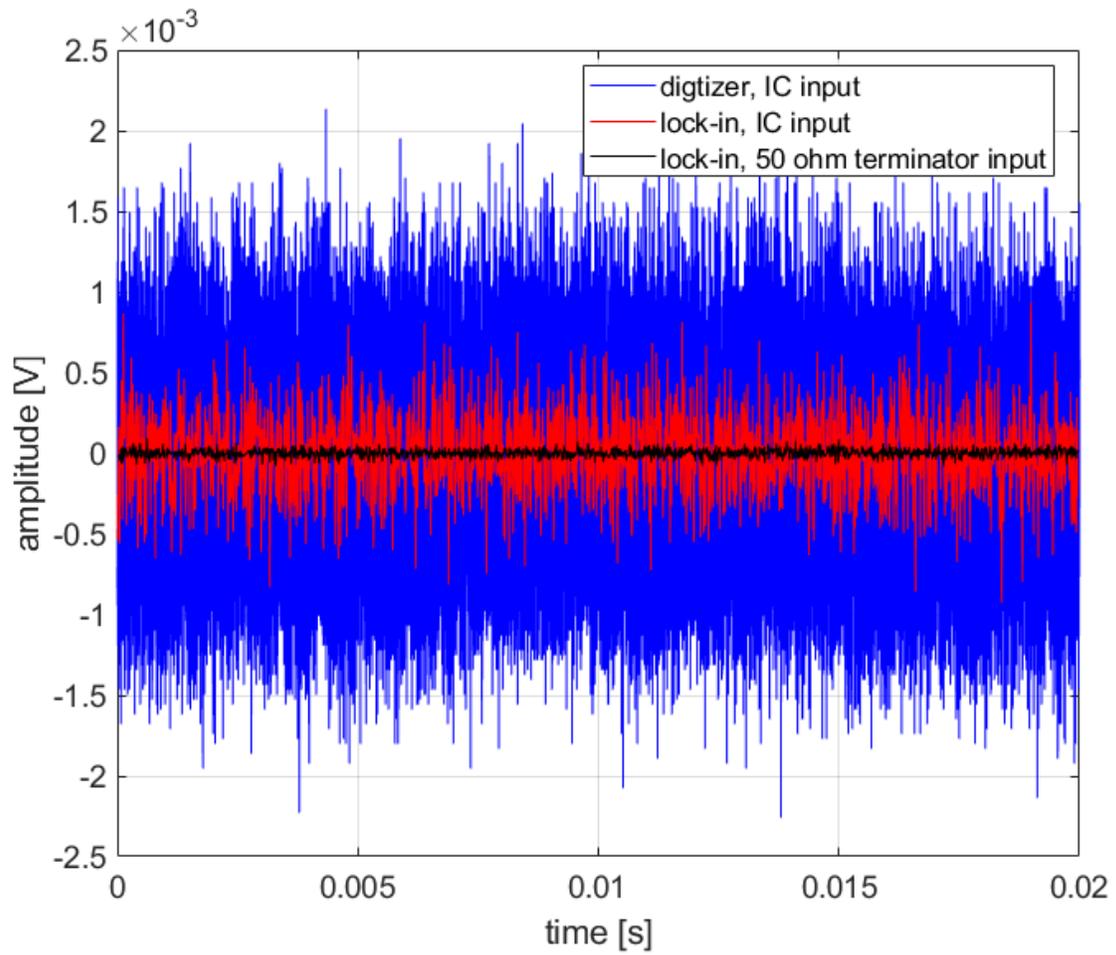

Figure 13 (SI): Background noise measurement of receive chain recorded with a digitizer (blue) and lock-in amplifier (red). The black trace is the lock-in measurement with 50 ohm termination.

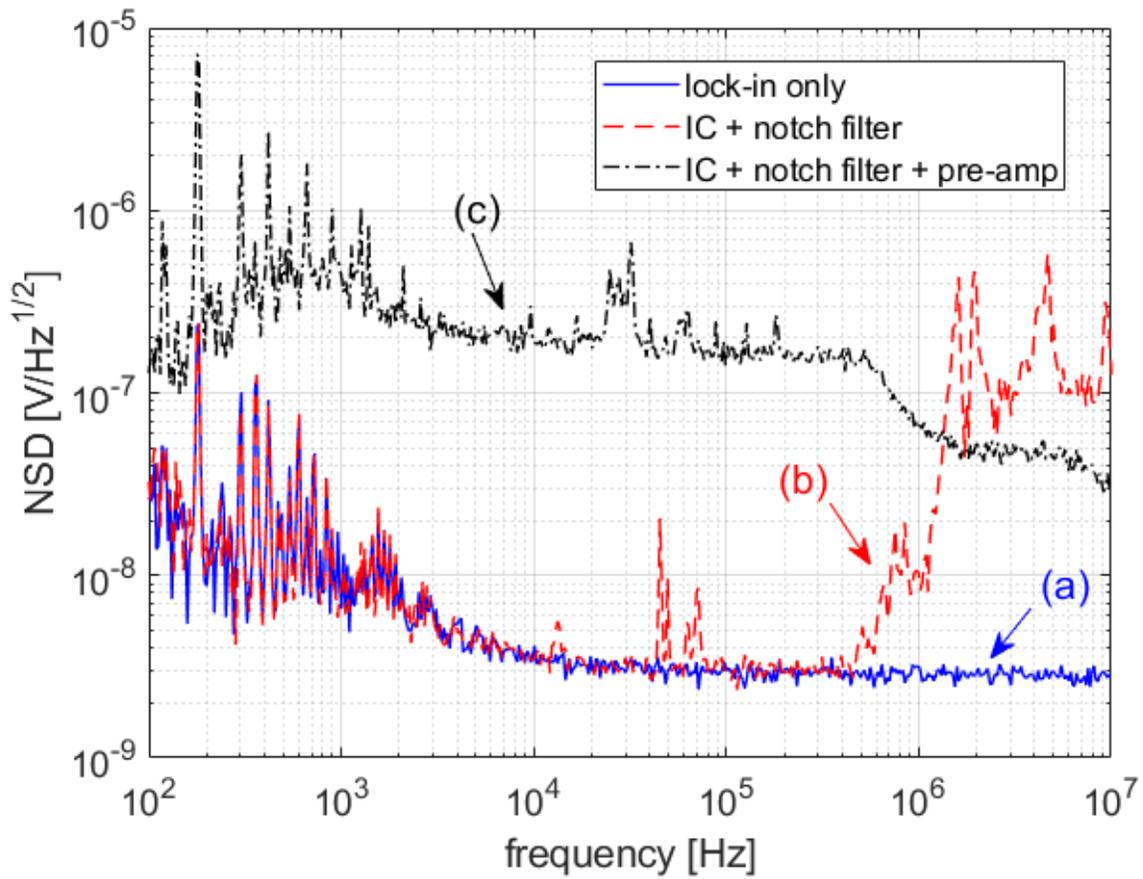

Figure 14 (SI): Noise spectral density of receive chain measured with a lock-in amplifier.

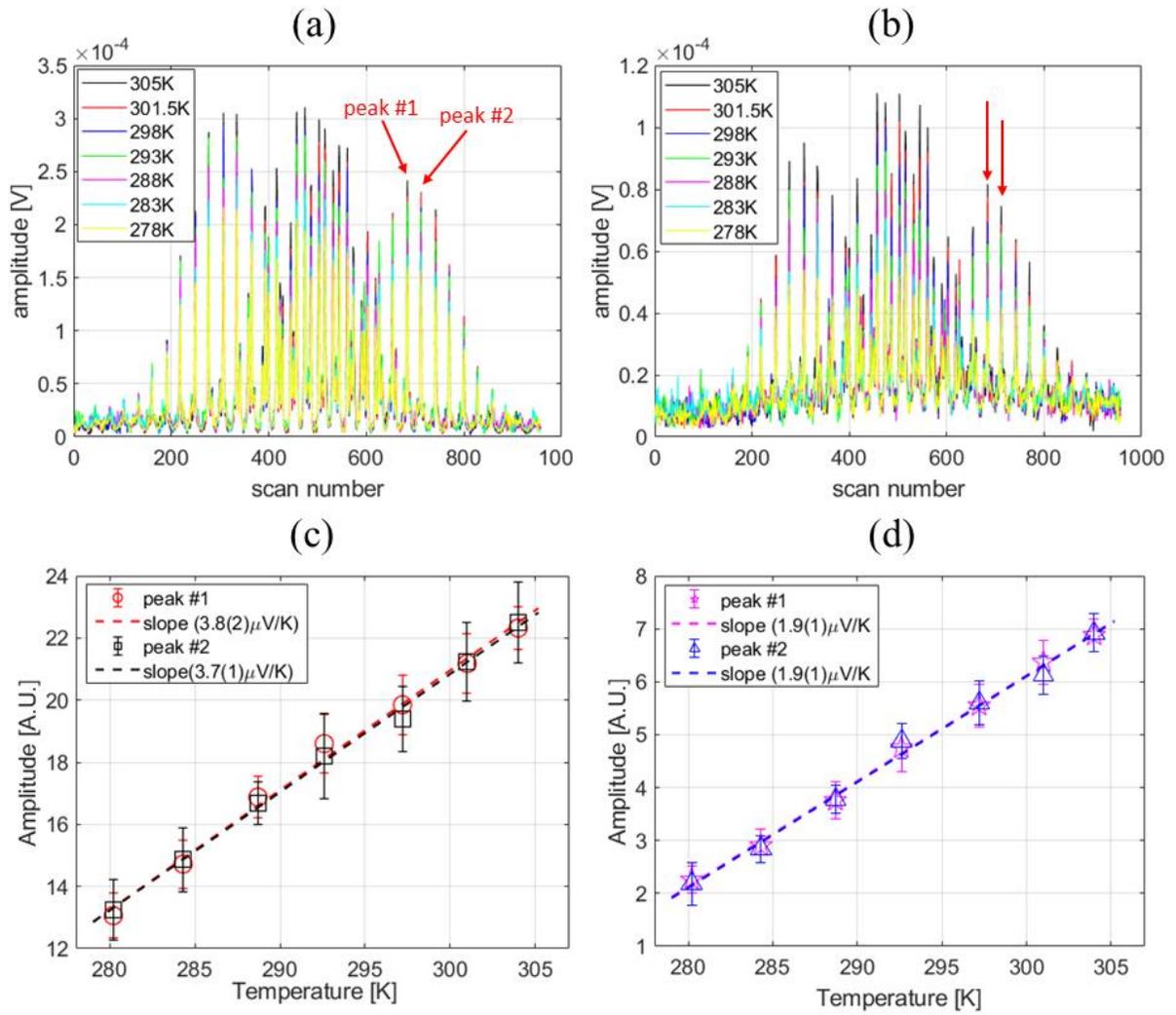

Figure 15 (SI): 1D representation of the 2D MPI image of Vivotrax in the glass phantom using the a) 7th and b) 9th harmonic. The two peaks indicated by red arrows were used to determine the temperature dependence of the c) 7th and d) 9th harmonic.

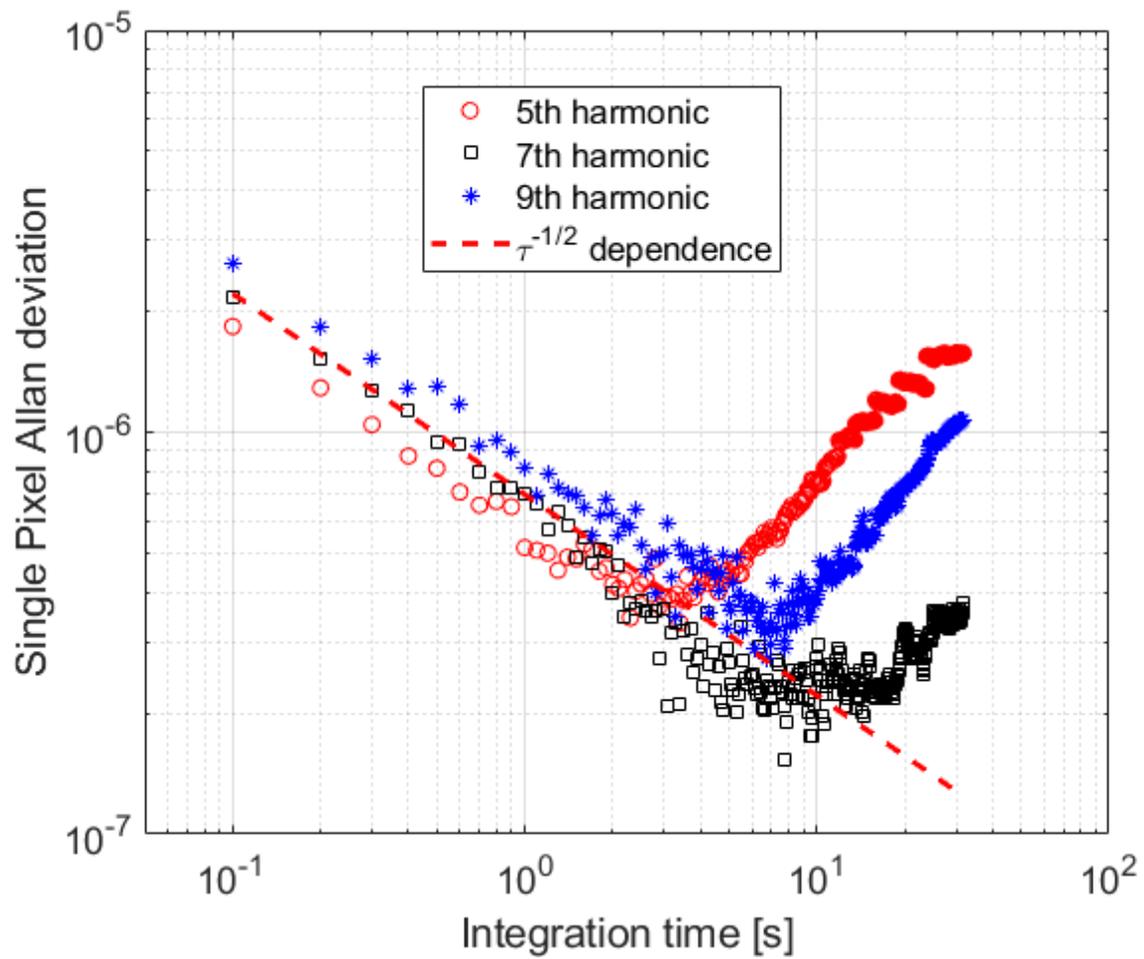

Figure 16 (SI): Allan deviation for harmonic MPI at a single pixel with Vivotrax tracers. The red dashed line shows the Allan deviation for a statistical (random) noise source.

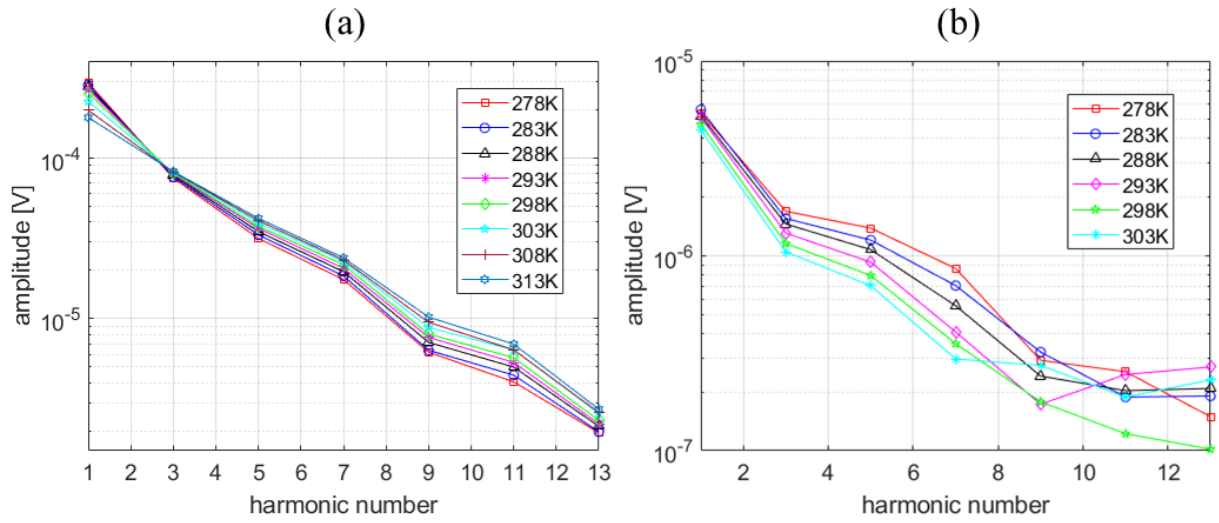

Figure 17 (SI): Temperature dependence of Vivotrax in a large glass vial using the (a) MPS and (b) MPI instrument.